\documentclass[epj]{svjour}
\usepackage{graphics}
\usepackage[demo]{graphicx}
\usepackage{amsmath}
\usepackage{float}
\begin{document}
\titlerunning{How spread changes affect the order book}
\title{How spread changes affect the order book: Comparing the price responses of order deletions and placements to trades}
\author{Stephan Grimm \and Thomas Guhr}                     
\offprints{}          
\institute{Fakult\"at f\"ur Physik, Universit\"at Duisburg-Essen, Lotharstra{\ss}e 1, 47048 Duisburg, Germany}
\date{Received: date / Revised version: date}
%
\abstract{We observe the effects of the three different events that cause spread changes in the order book, namely trades, deletions and placement of limit orders. By looking at the frequencies of the relative amounts of price changing events, we discover that deletions of orders open the bid-ask spread of a stock more often than trades do. We see that once the amount of spread changes due to deletions exceeds the amount of the ones due to trades, other observables in the order book change as well. We then look at how these spread changing events affect the prices of stocks, by means of the price response. We not only see that the self-response of stocks is positive for both spread changing trades and deletions and negative for order placements, but also cross-response to other stocks and therefore the market as a whole. In addition, the self-response function of spread-changing trades is similar to that of all trades. This leads to the conclusion that spread changing deletions and order placements have a similar effect on the order book and stock prices over time as trades.
\PACS{
      {PACS-key}{discribing text of that key}   \and
      {PACS-key}{discribing text of that key}
     } 
} 
\maketitle
\section{Introduction}
\label{intro}
The trading of stocks is central to and of vital importance in the activity of a stock market exchange. Those participating in the stock market often abide by the saying, “buy low and sell high”, simply in order to make the most profit on their trades within the market. Although often ignored in the model building, it has a large impact on the price dynamics and thus on the stylized facts as well as on the more specific features \cite{One,Two,Three,Four,Five,Six,Seven,Eight,Nine}. Furthermore, the relation between trades and price changes has received considerable attention \cite{Two,Four,Twelve,Thirteen,Fourteen,Fifteen,Sixteen,Seventeen,Eighteen}. Recent studies \cite{Eleven,Wang1,Wang2,Wang3,Bouchaud1} show that indeed trading of a stock not only has a positive effect on the traded stock’s price itself, but also on the prices of other stocks in the market as well. This indicates a more generalized impact on the stock market at large. However contrary to the naive assumption that trading is the predominant activity on the market, we will see in the sequel that the stock market is also severely affected by withdrawals of limit orders. In order to compare the effect of limit order deletions to that of trades, we look at events that cause a change in the spread, since trades can only happen at the quote prices of a stock. These events however include the placement of limit orders into the spread, therefore giving us a general overview of the effect of events that cause a spread change in the order book.

The paper is organized as follows: In Section \ref{Data set} we present our data set of stocks. We then analyze the different properties of spread changing events in section \ref{Properties}, first the distributions of relative amounts across stocks and then compare these to other observables in the order book. In section \ref{Self} we first introduce the sign of price change and the self-response function to then calculate the price response on a event and physical time scale. Finally in section \ref{Cross} we look at the individual cross-responses between stock pairs and the average market response to each event type. Our conclusion follows in section \ref{Conclusion}. Throughout this paper we will only analyze trades, deletions and placements of limit orders that cause a change in spread. For simplicity, each time we use the terms trades, deletions and order placement we imply that these are only events which change the spread of the stock.

\section{Data set}
\label{Data set}

The set of data we used comprises of 96 stocks of the NASDAQ100 stock index, as of February 21st, 2015. A detailed listing of all 96 stocks used can be found in appendix \ref{Appendix1}. For every stock we have the intra-day data of each entire day from the 7$^{th}$ to the 11$^{th}$ of March, 2016. This gives us 480 sets of daily stock data. We purposely selected a week without any unusually high price changes and the trading activity being about average for the respective stocks.
The format of the raw data set is the TotalView-ITCH format, which contains the order flow. This means that every change in the order book (an order being placed or exiting the order book) is listed with a unique identification number and a millisecond timestamp. In order to avoid overnight effects and any artifacts at the opening and closing of the market, we consider only trades of the same day from 9:40 a.m. to 3:50 p.m. New York local time.

\section{Properties of price changes}
\label{Properties}

\subsection{Distribution of price changes}
\label{Distribuion}

\begin{figure}[tbp]
\centering
\resizebox{0.5\textwidth}{!}{%
  \includegraphics{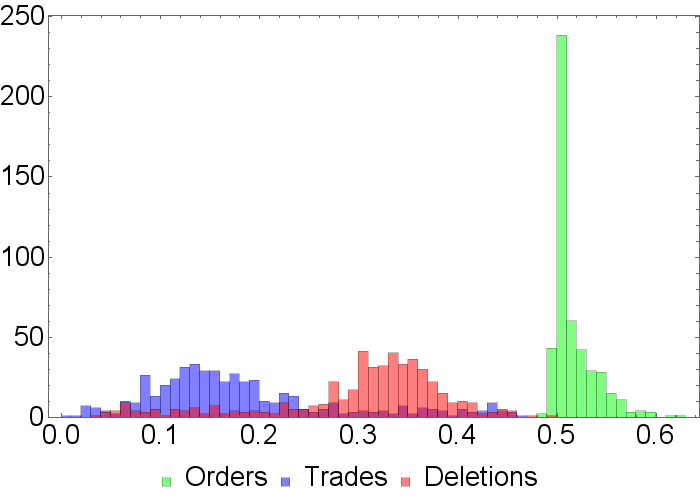}
}
\caption{Frequencies of spread changes due to order placements, trades and deletions across all 96 stocks for every day in the time period from the 7$^{th}$ to 11$^{th}$ of March, 2016. The ordinate shows the relative amount of spread changes due to each event type.}
\label{dist}       
\end{figure}

\begin{figure}[tbp]
\centering
\resizebox{0.5\textwidth}{!}{%
  \includegraphics{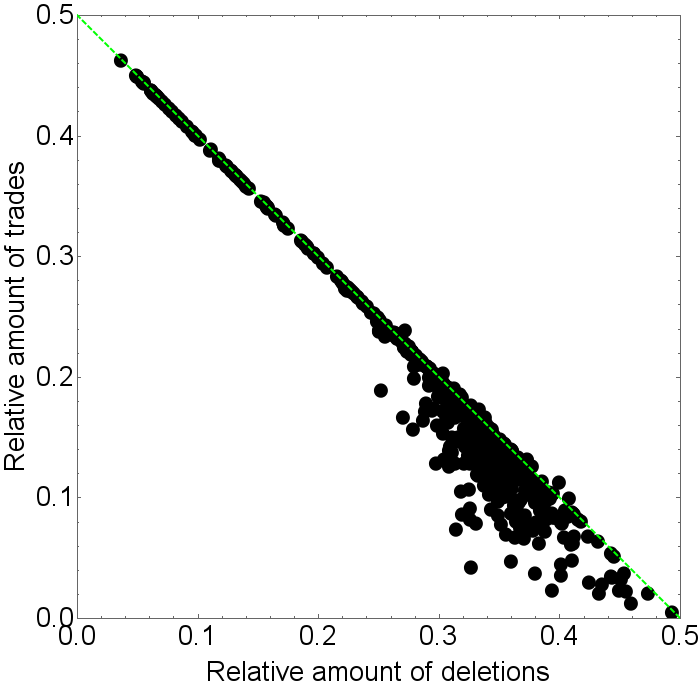}
}
\caption{Spread changes due to trades versus deletions shown as relative amounts. Each dot represents one day of the 96 stocks for every day from the 7$^{th}$ to 11$^{th}$ of March, 2016. The green dashed line represents the line of balance.} 
\label{tradel}       
\end{figure}

The price of a stock $i$ at a given time $t$ is given through the midpoint price $m_i(t)$. We use the same notations and conventions as in \cite{Wang2,Wang3}. The midpoint price can only change if best bid or best ask changes, so a midpoint price change always involves a change of the spread. The spread can widen through trade or deletions of orders and it can close by orders being placed into the spread. Figure \ref{dist} shows the histograms for all three types of spread changes with the respective relative amount occuring in each stock of our data set. We can observe three distinct distributions for trades, deletions and order placements. Most obvious is the very sharp peak for order placements, with the center of mass being around 0.5. The distributions for trades and deletions are much flatter, but each showing a noticable peak at different positions. The centers of those peaks are about 0.14 for trades and 0.34 for deletions. The distribution for deletions shows a heavy tail to the left of its peak, whereas the distribution for trades shows a heavy tail to the right. The spread seems to change most often due to placement of orders and least often by trades. Obviously the relationship
\begin{equation}
O + D + T = 1
\end{equation}
for spread changes is true for every individual stock, where as $O$, $D$ and $T$ are the relative amounts of spread changes due to order placements, deletions and trades, respectively. Because of the sharp peak in the distribution for order placements we approximate that the percentage of these is constant at 0.5. This yields the relation
\begin{equation} \label{inverse}
T \simeq  0.5 - D \quad .
\end{equation}
between the relative amounts of trades and deletions. Under the assumption of a balance of incoming and exiting orders on the spread level, we get the inverse relationship (\ref{inverse}) between the relative amounts of trades and deletions. Since the peak of the distribution for order placements is quite sharp, we see in Fig. \ref{tradel} this inverse relationship reflected in the empirical data when we plot the relative amounts of trades as function of the relative amounts of deletions. 

\begin{figure*}
\includegraphics[width=\linewidth]{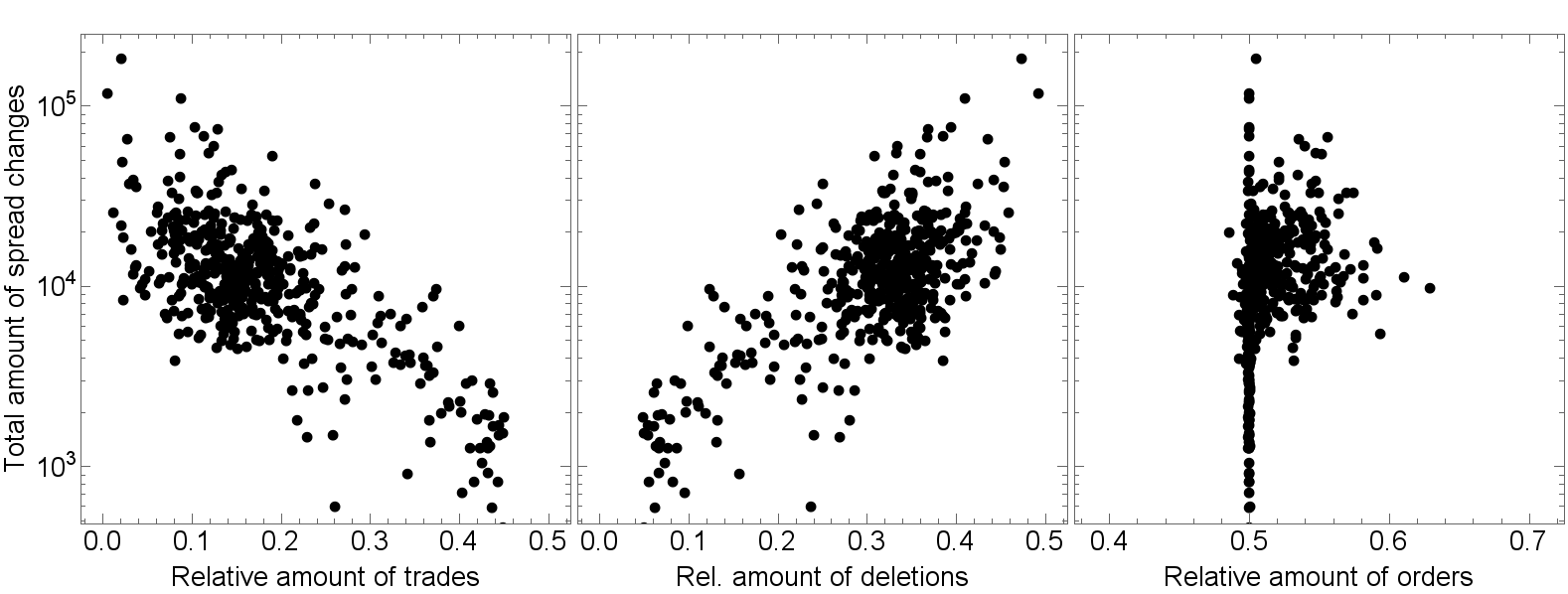}
\caption{Total amount of spread changes on a logarithmic scale versus the relative amounts of spread changes due to trades, deletions and order placements in a stock. Each dot represents one day of the 96 stocks for every day from the 7$^{th}$ to 11$^{th}$ of March, 2016.}
\label{alltot}       
\end{figure*}

However, the data points start to deviate more from the 0.5-line, which is shown as a green dashed line, the higher the relative amount of deletions gets. We will refer to the green line as the \textit{line of balance}, as it marks the point where the sum of spread changes due to deletions and trades is equal to the amount of orders placed into the spread, thus causing a balance of orderflow on the spread level. The deviation only appears to happen below the green line, as very few data points are found above it. When a data point is below the line of balance, it means that trades and deletions together make up less than half of the events that change the spread in that particular stock. Consequentially this means that more orders are being placed into the spread than the spread being changed through trade or deletion. This corresponds well with the distribution for order placements, as it has a noticable tail to the right of its peak but a very light tail to the left. This means there are more stocks in which the relative amount of order placements at the spread level is above half. Interestingly enough the deviations only appear to happen once a stock has its spread changes caused by deletions more than 0.25 of the time or in other words when the spread is more often changed due to a deletion than a trade. On the other hand, if we have more trades than deletions on the spread level, there always seems to be the aforementioned balance of incoming and exiting orders.

If one equates the amount of spread changes caused by exiting orders with the activity in the orderbook, Fig. \ref{alltot} indicates that activity mostly is made up by deletions. As the activity in a stock increases, spread changes due to trades become less likely. A possible explanation for this is that the more spread changes there are in a stock, the less time there is for volume to be accumulated on the quotes. In other words, there is less time for limit orders to be put into the order book at the same price as the best bid and best ask. Therefore, less individual orders are at the spread level. As there are less orders on the quote, it is more likely that if an order at the spread level is deleted that it is the last remaining order there and therefore changes the spread. The difference between trades and deletions is that deletions can only affect one order as opposed to trades being able to affect an order either fully, partially or multiple orders at once, depending on the volume of the initiating order. Trades can thus also change the spread even if there is more than one order on the spread level. 

A relation between the amount of activity and the relative amount of order placements only seems to visible at both ends of the activity spectrum. These stocks are then located on the line of balance, as all of the data points are located at a relative amount of order placements of 0.5.

\subsection{Spread changes vs. other observables}
\label{Observables}
\begin{figure*}
\includegraphics[width=\linewidth]{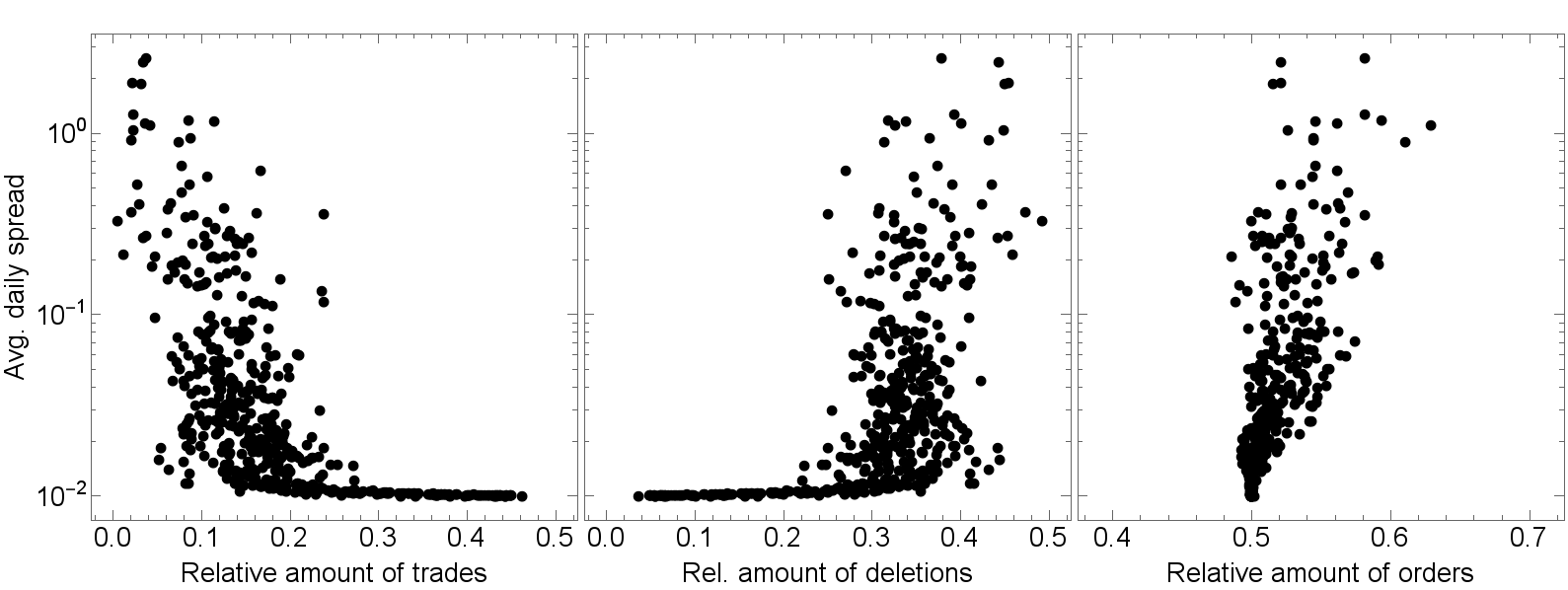}
\caption{Average time-weighted daily spread on a logarithmic scale versus the relative amounts of spread changes due to trades, deletions and order placements in a stock. Each dot represents one day of the 96 stocks for every day from the 7$^{th}$ to 11$^{th}$ of March, 2016.}
\label{allspread}       
\end{figure*} 

\begin{figure}[htbp]
\centering
\resizebox{0.5\textwidth}{!}{%
  \includegraphics{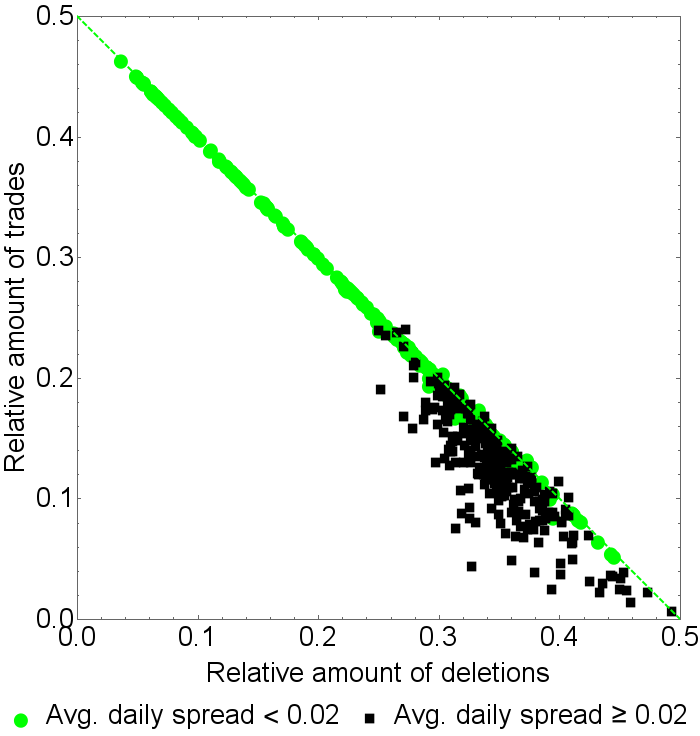}
}
\caption{Spread changes due to trades versus deletions shown as relative amounts. Green dots indicate stocks on different days which have an average daily spread of below 0.02. Each dot represents one day of the 96 stocks for every day from the 7$^{th}$ to 11$^{th}$ of March, 2016. The green dashed line represents the line of balance.}
\label{tradel2}       
\end{figure}

Figure \ref{allspread} shows the relation between the distribution of the three event types and the average daily spread. The average spread in our case is the arithmetic mean, weighted with the amount of time the particular quote remains in the orderbook. 
As the spread size increases, the relative amount of order placements starts to deviate more from 0.5. High relative amounts of trades and low relative amounts of deletions in a stock coincide with narrow average spreads, which is a direct result of the inverse relationship between trade and deletion percentages as shown in Fig. \ref{tradel}. The inverse relationship also is reflected in the seemingly mirrored shapes of the two graphs. As before, we again have two disctinct regimes on either side of the 0.25-line, although for both deletions and trades. In the regime when trades exceed the amount of deletions we only find narrow spreads which are all slightly above the minimum possible width of an average spread, which is 0.01. Stocks with an average spread close to this value remain at this spread for most of the day, which impacts the time-weighted average. In the regime where the amount of deletions exceeds the amount of trades, we find a wide variety of spreads. It appears as if a relationship does not exist between the percentage of deletions or trades and the width of the average spread. If we compare Fig. \ref{allspread} with Fig. \ref{tradel}, more specifically the regime in which the amount of trades exceeds the amount of deletions, we see that for this regime stocks that have a narrow average spread close to the minimum are right on the line of balance. This means that if a spread of 0.01 opens up, the next step is an order being placed into the spread, thus closing it again. This process repeats throughout the day, therefore causing order placements to be the same amount as orders exiting the order book. In Fig. \ref{tradel2} the stocks with an average spread below 0.02 are highlighted in green. They are all located on the line of balance. So if a stock has an average spread close to the miminum, there generally will be the same amount of orders being placed into the spread as orders exiting the orderbook. This is because in those stocks the price levels close to the spread are all occupied with many orders and therefore the spread usually only changes by 0.01. The changes happen between the two smallest possible values 0.01 and 0.02, with the spread alternating between opening and closing. However this balance does not only appear in these particular stocks but also in stocks with a larger average spread as we can also see multiple black dots which are located on the line of balance.

\section{Average self-response}
\label{Self}

\subsection{Sign of price change}
\label{Sign}

We introduce the \textit{sign of price change} in stock $i$ for an event at the time $t$ as
\begin{equation}
\varepsilon_i (t) = \textrm{sgn} \Big( S_i(t) - m_i(t-\delta) \Big)=
\begin{cases}
      +1 & \text{if on ask side,} \\
      -1 & \text{if on bid side,}
    \end{cases}
\end{equation}
where $S_i(t)$ is the price of the limit order which is placed into or exits the order book at the time $t$ and $m_i(t-\delta)$ the midpoint price immediatly before the event takes place.

\subsection{Price response function}
\label{Function}

In order to measure the effect of a spread changing event on the stock price of a stock $i$ over time, we use the logarithmic return, defined as
\begin{equation}
r_i(t,\tau) = \log m_i(t+\tau) - \log m_i(t-\delta) = \log \frac{m_i(t+\tau)}{m_i(t-\delta)} \quad .
\end{equation}
To acquire statistical significance, the price response function \cite{Eleven,Wang2}
 \begin{equation}
 R_i(\tau) = \Big \langle \varepsilon_i(t) r_i(t,\tau) \Big \rangle_t
 \end{equation}
 is the time average of the product of time-lagged returns and trade signs for the stock $i$.

\begin{figure}[htbp]
\centering
\resizebox{0.48\textwidth}{!}{%
  \includegraphics{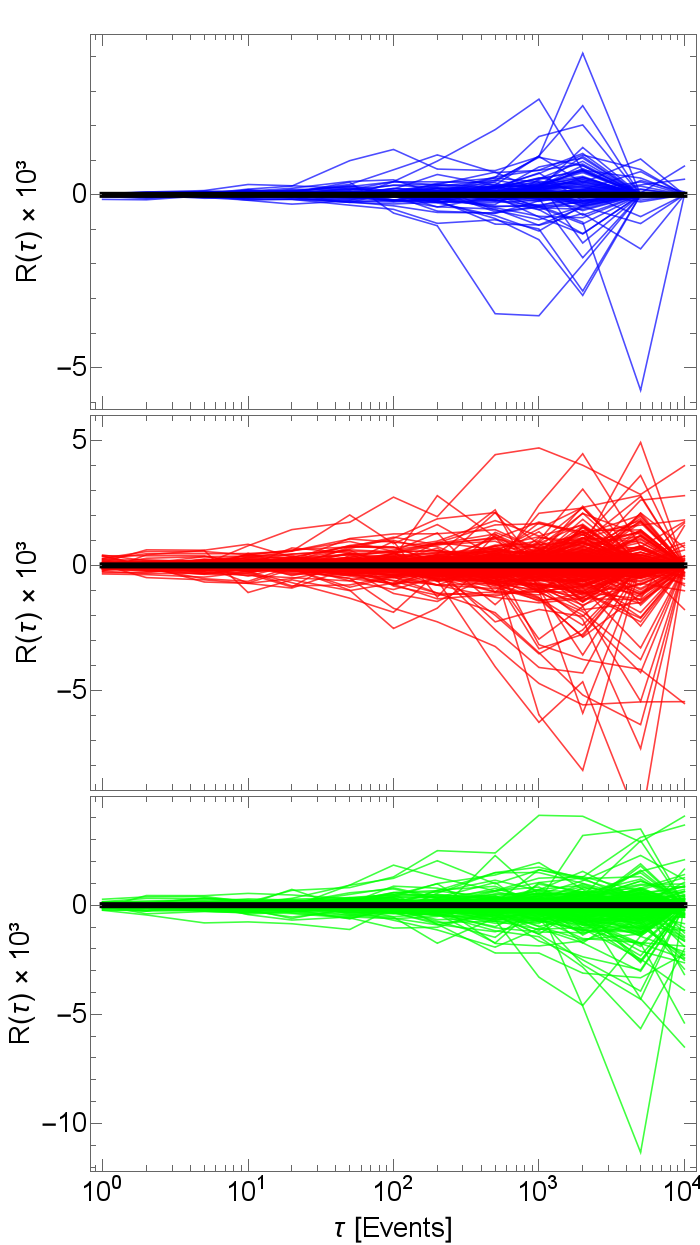}
}

\caption{Price response on the logarithmic event time scale for spread changes due to trades (top), deletions (middle) and order placements (bottom). The black line is the average of all the response curves. Each line represents one day of the 96 stocks for every day from the 7$^{th}$ to 11$^{th}$ of March,
2016.}
\label{evall}       
\end{figure}

\subsection{Price response on event time scale}
\label{Event}

\begin{figure}[htbp]
\centering
\resizebox{0.48\textwidth}{!}{%
  \includegraphics{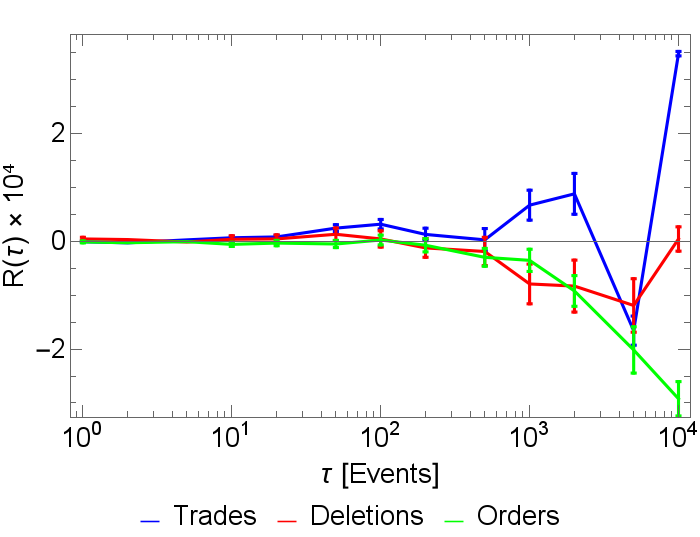}
}
\caption{Average price response on the logarithmic event time scale for spread changes due to trades, deletions and order placements. Each line represents the average over each of the 96 stocks for every day from the 7$^{th}$ to 11$^{th}$ of March, 2016.}
\label{evmean}       
\end{figure}

We calculate the response function of all 96 stocks across each of the five days, shown in Fig. \ref{evall}. To do so, we measure the time on the event scale which means that every time step is another event of the same type. Here we only calculate intra-day price responses which means that if the time $t+\tau$ is outside of the data for that day, the response for that event is not included in the calculation. The response functions are calculated for a time lag of up to 10000 events. We see that the graphs for all three events are similar and in the same order of magnitude. To better compare the effect of the three event types we plot the average responses in Fig. \ref{evmean}. Trades show an increasing positive price response, with the function returning to zero after about 5000 events, indicating a movement of the price in the direction of the trade. The price response to order placements is increasingly negative and the price response to deletions starts off positive but than falls below zero after about 100 events. After its negative peak it returns back to zero after about 10000 events, indicating a longer lasting effect than trades, but shorter than order placements. The large jump in the response functions of trades at 10000 seconds is likely due to a lack of available data for that time lag. 

\begin{figure}[htbp]
\centering
\resizebox{0.48\textwidth}{!}{%
  \includegraphics{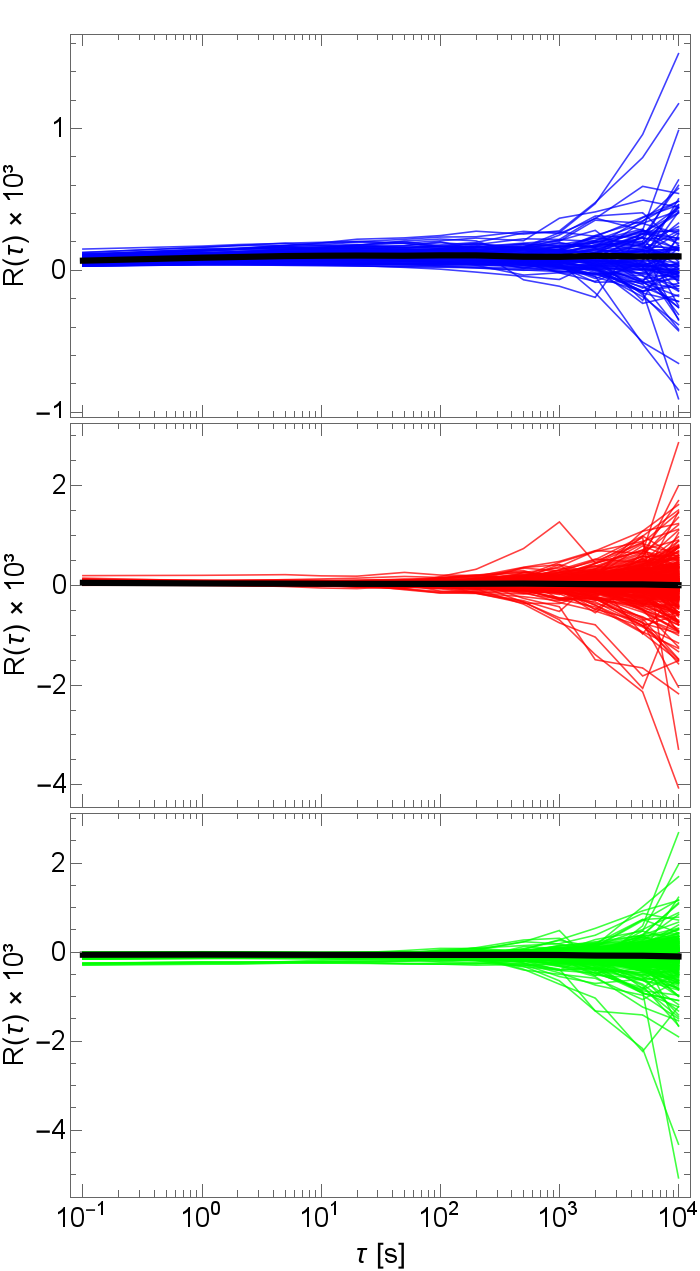}
}
\caption{Price response on the logarithmic physical time scale in seconds for spread changes due to trades (top), deletions (middle) and order placements (bottom). The black line is the average of all the response curves. Each line represents one day of the 96 stocks for every day from the 7$^{th}$ to 11$^{th}$ of March,
2016.}
\label{allself}       
\end{figure}

\subsection{Price response on physical time scale}
\label{Physical}

\begin{figure}[htbp]
\centering
\resizebox{0.48\textwidth}{!}{%
  \includegraphics{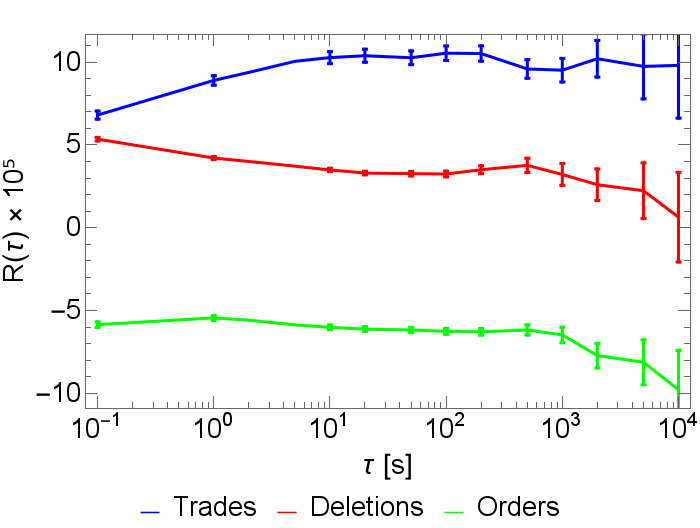}
}
\caption{Average price response on the logarithmic physical time scale in seconds for spread changes due to trades, deletions and order placements. Each line represents the average over each of the 96 stocks for every day from the 7$^{th}$ to 11$^{th}$ of March, 2016.}
\label{selfmean}       
\end{figure}

The time between two events is not constant and not only does the average over all 96 stocks across the five days produce different physical time scales, since the next event can happen at different time, but it also only measures the spread change compared to the next event. The response on an event time scale does not register what happens with the price in between events of the same type since there can be other events causing the price to change. In order to see how the price is immediately affected by an event, we now calculate the price response of a spread changing event on a physical time scale, with the results shown in Fig. \ref{allself}. The response functions are calculated for a time lag of up to 10000 seconds. Again, beyond 1000 seconds or so, the results lose their statistical significance. As for the response function on an event time scale, the individual responses for each of the three event types have the same shape. To better compare the effect of the three event types we plot the average responses in Fig. \ref{selfmean}. All three averages for each event type show a jump immediately after the spread change takes place, which is result of the return definition. Both the response of trades and order placements is in the same direction as before on the event time scale. Deletions however show a positve price response on a physical time scale. Noticable is also that the price response of deletions decreases after the initial jump, as opposed to trades where the response function increases and then reverses.

The responses for trades and order placements are both larger in absolute value than that of deletions. This can be explained by the ability of trades and order placements to cover multiple tick levels, as opposed to deletions that are limited to one tick level, of course assuming that the price levels outside of the spread are filled with limit orders. In total the response on a physical time scale is about one magnitude larger than that on an event time scale.

\subsection{Quote changing events vs. all events in the order book}
\label{Full}

\begin{figure}[htbp]
\centering
\resizebox{0.48\textwidth}{!}{%
  \includegraphics{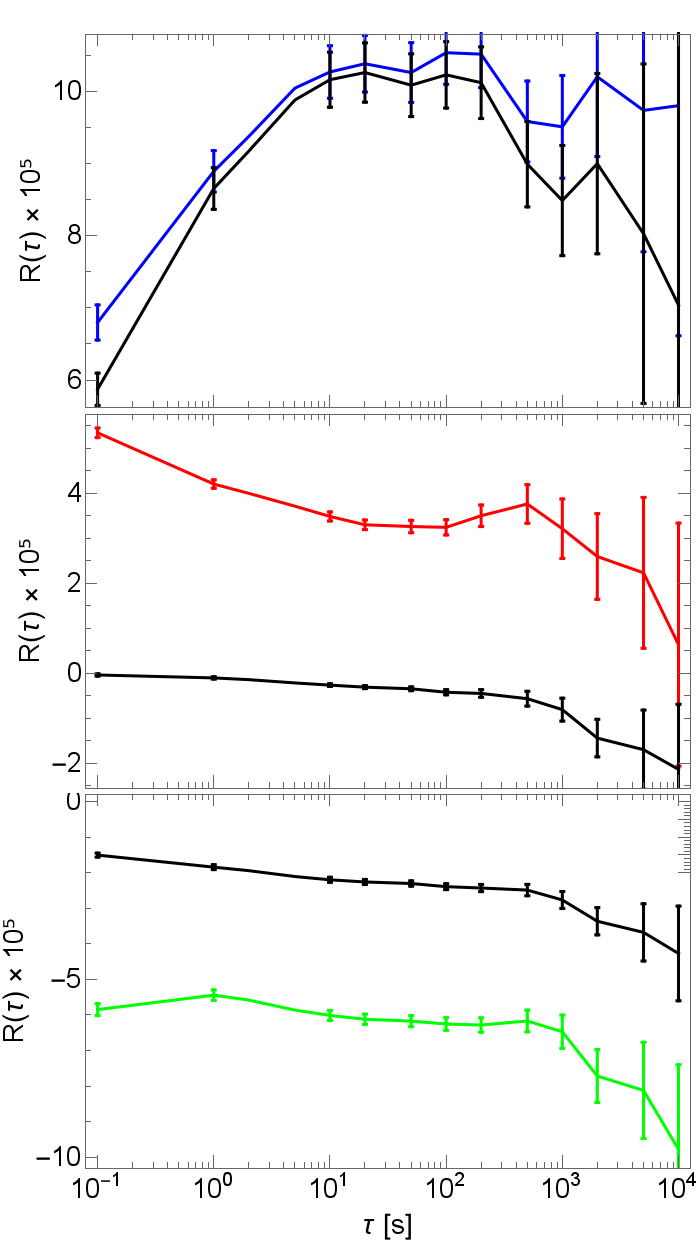}
}
\caption{Price response on the logarithmic physical time scale in seconds for spread changes due to trades (top), deletions (middle) and order placements (bottom). The colored line is the average response for corresponding events that change the spread and the black line is the response for all corresponding events. Each line represents the average over each of the 96 stocks for every day from the 7$^{th}$ to 11$^{th}$ of March, 2016. We notice the different abscissae.}
\label{fullself}       
\end{figure}

In Fig. \ref{fullself} we compare the price responses to events that change the spread to the responses of all events of that type in the order book, regardless if they change the midpoint price or not. The two response curves for trades show remarkable similarity with only a small vertical offset between the two, as the response to all trades in the order book is slightly smaller. This can be explained by the fact that the trades that do not change the midpoint price directly do not exhibit a price jump at $\tau = 0 s$. This leads to the conclusion that trades that do not change the spread have a smaller price response than the ones that do, albeit that the difference between the two is quite small. Therefore in order to measure the price response of trades in a stock, it is sufficient to look at quote changing trades. Contrary to this result, the price responses to deletions and order placements are vastly different if we only take into account the events that change the spread as opposed to all of them. In both cases the response curve for all events is close to zero, indicating that those events as a whole do not effect the price. This makes sense, because on the one hand limit orders are being placed and deleted in large numbers on both sides of the spread and on the other hand happen not only at the spread level, but also deep inside the order book. Both of these facts lead to the events on both sides of the spread cancelling each other out and causing the price response to drop to zero. That the averages over all events for both deletions and order placements move away from zero for time lags $\tau$ greater than 1000 seconds, is due to the fact that they lose their statistical significance. Contrary to trades, deletions and order placements that change the spread seem to carry more information in regards to the price response than those that do not.

\section{Average cross-response}
\label{Cross}

\subsection{Cross-response for individual stock pairs}
\label{Cross pairs}

\begin{figure*}[htbp]
\includegraphics[width=\linewidth]{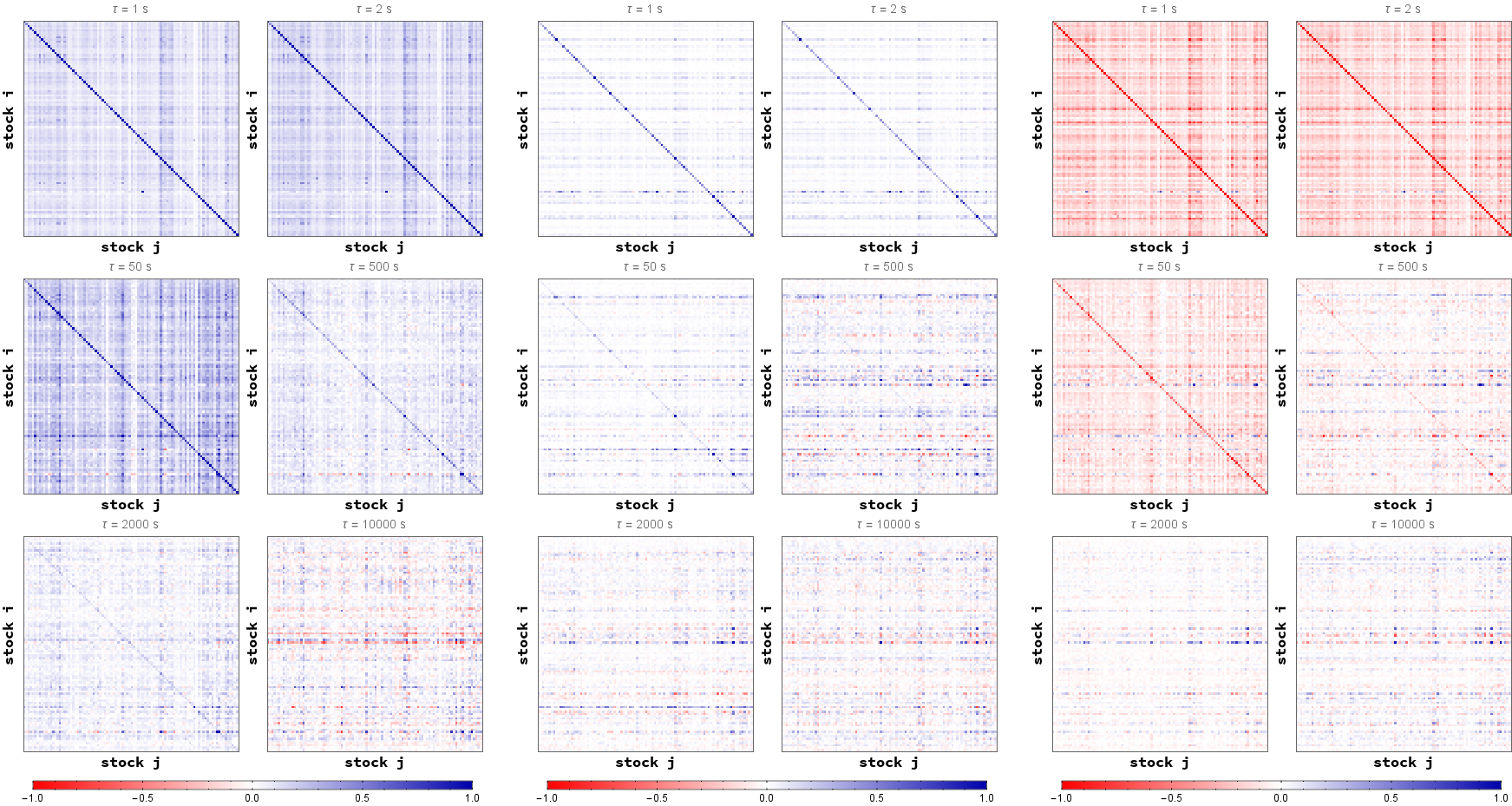}
\caption{Standardized cross-response matrix with $\tau$ = 1,2,50,500,2000,10000 seconds for spread changes due to trades (left), deletions (midle) and order placements (right). Each row $i$/column $j$ represents one of the 96 stocks averaged over every day from the 7$^{th}$ to 11$^{th}$ of March, 2016.}
\label{crall}       
\end{figure*}

To calculate the effect of spread changing events in one stock on the prices of other stocks we introduce the cross-response function \cite{Wang2}
\begin{equation}
 R_{ij}(\tau) = \Big \langle \varepsilon_i(t) r_j(t,\tau) \Big \rangle_t \quad ,
 \end{equation}
as the average over all spread changes in the stock $i$ multiplied with the time-lagged returns in the stock $j$. When $i = j$ this function gives us the self-response we discussed in section \ref{Physical}. For the cross-response we use the physical time scale, because there is no reason to assume that events in different stocks are synchronized. If we average the cross-response for every stock pair $i,j$ over all five days for a fixed value of $\tau$ we get a $96 \times 96$-Matrix $\rho(\tau)$. These matrices are shown in Fig. \ref{crall} for different time lags $\tau$ = 1,2,50,500,2000,10000 seconds. Rows $i$ indicate the stocks in which the spread changing events occur. Columns $j$ indicate the stocks in which the spread change is measured after events occur in other stocks. The stocks are in alphabetical order as shown in appendix \ref{Appendix1}. For better visualization we normalize each matrix element 
\begin{equation}
\rho_{ij}(\tau) = \frac{R_{ij}(\tau)}{\max(R_{i\ne j}(\tau))}
 \end{equation}
to the largest absolute off-diagonal value of the corresponding cross-response matrix \cite{Wang2,Wang3}. The color scale, as shown to the right of each pair of matrices, ranges from red, indicating a value of −1, to blue, indicating a value of +1. The cross responses for each of the three different event types show a visible diagonal line for lower values of $\tau$, which means that in general the self-response is larger at first than the cross-response between stocks, but slowly blends in with the cross-responses for all three event types, as $\tau$ increases. Furthermore all matrices display clear patterns of strips that are quite stable over time. The cross-responses for trades is mostly positive, with only a few strips being red. The cross-response matrices for order placements are the opposite; mostly positive, with only a few strips being blue. For deletions both blue and red strips appear in equal amounts. The appearance of strips indicates, that there are certain stocks that strongly influence the entire market (active response) and stocks that are influenced strongly by all the other stocks (passive response), which are defined as
\begin{equation}
R_j^{(a)}(\tau) = \langle R_{ij}(\tau) \rangle_i \quad \textrm{and} \quad R_i^{(p)}(\tau) = \langle R_{ij}(\tau) \rangle_j ,
\end{equation}
respectively. The cross-response matrices for trades get darker at first and then lighter as $\tau$ increases. For order placements the cross-response matrix is the most colorful at first and then gets whiter over time. The cross-response for deletions does not change its intensity much over time, as it stays lighter for all values of $\tau$.

\subsection{Market response}
\label{Market}

\begin{figure}[htbp]
\centering
\resizebox{0.48\textwidth}{!}{%
  \includegraphics{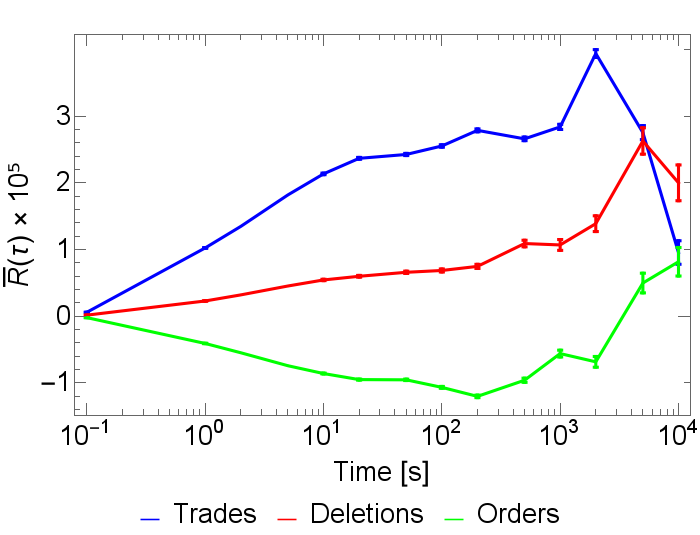}
}
\caption{Market Response (Averages of the cross responses) on the physical time scale in seconds for spread changes due to trades, deletions and order placements. Each line represents the average of the 96 stocks of every day from the 7$^{th}$ to 11$^{th}$ of March, 2016.}
\label{crmean}       
\end{figure}

Since the cross-response matrices are a snapshot in time of the average price response of stocks to certain events in other stocks they give us information of the general effect trades, deletions and order placements have on the market as a whole. In order to measure these price impacts and acquire statistical significance we calculate the market response by averaging over every off-diagonal element of every matrix and therefore getting the average crossresponse as a function of the time-lag $\tau$ \cite{Wang2,Wang3}
\begin{equation}
\overline{R}(\tau) = \Big \langle \big \langle R_{ij}(\tau) \big \rangle_i \Big \rangle_j \quad \text{with $i\neq j$}.
 \end{equation}
Figure \ref{crmean} shows that the average market response is positive for both trades and deletions, and negative for order placements. This is in line with the results for the self-response on a physical time scale in section \ref{Self}. Both functions for trades and deletions show a major peak at 2000 and 5000 seconds respectively before reverting back to zero. Before their major peak both functions exhibit a smaller peak. The average market-response function for order placements is negative with a major peak at 500 seconds and a minor peak after 2000 seconds, before reverting back to zero. The two-peak structure for all three curves is an effect of averaging over both rows and columns in the matrices, which correspond to active and passive cross-responses. When averaging over the entire market we see a long lasting effect (10$^{4}$ seconds) on prices to all three event types. Due to the average over both indices $i$ and $j$, the results gain a high statistical significance, also for long time scales beyond 1000 seconds. These results show the same behaviour as \cite{Wang3}.
 
\section{Conclusion}
\label{Conclusion}

We have looked at events that change the price of a stock. Three events can change the price: An order being placed into the spread and a trade or a deletion of a limit order that changes the quote. We have seen that the price in most stocks is more often changed by deletions of orders than trades and that order placements make up about half of the spread changes and that the ratio of trades to deletions in a stock can have an effect on other observables in the orderbook. Once the amount of trades exceeds the amount of deletions the average daily spread seems to be close to the minimal possible amount. Stocks with this property generally show a balance in spread changes caused by order placements and orders exiting the orderbook. To compare the effect of spread changes caused by trades, deletions and order placements on the stock price over time, we calculated the self-response on both an event and a physical time scale. In both cases the price response to trades was positive and negative to order placements. As they were both of similar shape and magnitude, albeit opposite signs, both event types seem to have a comparable effect on the price of the stock they occured in. For deletions however the response on an event time scale was negative and positive on a physical time scale. In both cases the response was smaller in value than trades or order placements, indicating a lesser impact on prices due to deletions. This however could just be an effect of deletions only being able to change a quote by one tick, as opposed to trades and order placements that can change quotes by multiple ticks at once. Comparing the price responses of the events that change the spread to the price responses of all events, we saw that trades show a remarkable similarity for both cases, leading to the conclusion that the price response to trades can be calculated using only trades that change the price. However, for deletions and order placements the price responses for all events were close to zero, as opposed to the reponse of events that change the price. This shows that deletions and order placements that do change the spread do carry some kind of additional information over the other deletions and order placements in the order book. We then looked at the effect of spread changing events on the prices of other stocks, by means of the cross-response function. As for the self-response, the cross-response matrices for trades and order placements appeared similar when taking into account the sign of spread change. The cross-response matrices for both these event types showed a strong effect in the beginning, indicating a strong effect spanning across the entire market. For deletions the matrices were mostly white in color, indicating a rather small effect on the market. To better visualize the impact of the spread changing events we looked at the average market response functions. As the average is ober both the stock indices $i$ and $j$ (with $i \neq j$), the result is statistically significant for longer time scales beyond 1000 seconds. All three event types exhibit a non-zero price response across the market with a long term effect that lasts about 10$^{4}$ seconds, with the response to trades and deletions being positive and negative to order placements. As with the self-response the market-response to deletions is lower in value than that to trades or order placements. This leads us to the conclusion that deletions and order placements have a comparable effect on the dynamic in the order book, at least when looking at spread changes due to these events.

	\bibliography{Paper}
\bibliographystyle{ieeetr}

\newpage
\appendix

\section{Data set} \label{Appendix1}

The following table shows a detailed list of all the stocks used in this paper.

\begin{table}[H]
\label{table}       
\caption{List of 96 NASDAQ100 stocks}

\begin{tabular}{lll}
\hline\noalign{\smallskip}
Symbol & Name \\
\noalign{\smallskip}\hline\noalign{\smallskip}
AAL & American Airlines Group, Inc. \\
AAPL & Apple Inc. \\
ADBE & Adobe Systems Incorporated \\
ADI & Analog Devices, Inc. \\
ADP & Automatic Data Processing, Inc. \\
ADSK & Autodesk, Inc. \\
AKAM & Akamai Technologies, Inc. \\
ALXN & Alexion Pharmaceuticals, Inc.\\
AMAT & Applied Materials, Inc.\\
AMGN & Amgen Inc.\\
AMZN & Amazon.com, Inc.\\
ATVI & Activision Blizzard, Inc\\
AVGO & Broadcom Limited\\
BBBY & Bed Bath \& Beyond Inc. \\
BIDU & Baidu, Inc.\\
BIIB & Biogen Inc.\\
BMRN & BioMarin Pharmaceutical Inc.\\
CA & CA Inc.\\
CELG & Celgene Corporation\\
CERN & Cerner Corporation\\
CHKP & Check Point Software Tech. Ltd.\\
CHRW & C.H. Robinson Worldwide, Inc.\\
CHTR & Charter Communications, Inc.\\
CMCSA & Comcast Corporation\\
COST & Costco Wholesale Corporation\\
CSCO & Cisco Systems, Inc.\\
CTSH & Cognizant Technology Solutions Corp.\\
CTXS & Citrix Systems, Inc.\\
DISCA & Discovery Communications, Inc.\\
DISH & DISH Network Corporation\\
DLTR & Dollar Tree, Inc.\\
EA & Electronic Arts Inc.\\
EBAY& eBay Inc.\\
EQIX & Equinix, Inc.\\
ESRX & Express Scripts Holding Company\\
EXPD & Expeditors International of WA, Inc.\\
FAST & Fastenal Company\\
FB& Facebook, Inc.\\
FISV& Fiserv, Inc.\\
FOXA & Twenty-First Century Fox, Inc.\\
GILD & Gilead Sciences, Inc.\\
GOOG & Alphabet Inc.\\
GRMN & Garmin Ltd.\\ 
\noalign{\smallskip}\hline
\end{tabular}
\end{table}

\begin{table}[H]
\begin{tabular}{lll}
\hline\noalign{\smallskip}
Symbol & Name \\
\noalign{\smallskip}\hline\noalign{\smallskip}
HSIC & Henry Schein, Inc.\\
ILMN & Illumina, Inc.\\
INTC & Intel Corporation\\
INTU & Intuit Inc.\\
ISRG & Intuitive Surgical, Inc.\\
JD & JD.com, Inc.\\
KHC & The Kraft Heinz Company\\
KLAC & KLA-Tencor Corporation\\
LBTYA & Liberty Global plc\\
LLTC & Linear Technology Corporation\\
LMCA & Liberty Media Corporation\\
LRCX & Lam Research Corporation\\
LVNTA& Liberty Interactive Corporation\\
MAR & Marriott International\\
MAT & Mattel, Inc.\\
MDLZ & Mondelez International, Inc.\\
MNST & Monster Beverage Corporation\\
MSFT & Microsoft Corporation\\
MU& Micron Technology, Inc.\\
MYL & Mylan N.V.\\
NFLX & Netflix, Inc.\\
NTAP & NetApp, Inc.\\
NVDA & NVIDIA Corporation\\
NXPI & NXP Semiconductors N.V.\\
ORLY & O’Reilly Automotive, Inc.\\
PAYX & Paychex, Inc.\\
PCAR & PACCAR Inc.\\
PCLN & The Priceline Group Inc.\\
QCOM & QUALCOMM Incorporated\\
REGN & Regeneron Pharmaceuticals, Inc.\\
ROST & Ross Stores, Inc.\\
SBAC & SBA Communications Corporation\\
SBUX & Starbucks Corporation\\
SIRI & Sirius XM Holdings Inc.\\
SNDK& SanDisk Corporation\\
SPLS & Staples, Inc.\\
SRCL & Stericycle, Inc.\\
STX & Seagate Technology PLC\\
SYMC & Symantec Corporation\\
TRIP & TripAdvisor, Inc.\\
TSCO & Tractor Supply Company\\
TSLA & Tesla Motors, Inc.\\
TXN & Texas Instruments Incorporated\\
VIAB & Viacom Inc.\\
VIP & VimpelCom Ltd.\\
VOD & Vodafone Group Plc\\
VRSK & Verisk Analytics, Inc.\\
VRTX & Vertex Pharmaceuticals Incorporated\\
WDC & Western Digital Corporation\\
WFM & Whole Foods Market, Inc.\\
WYNN & Wynn Resorts, Limited\\
XLNX & Xilinx, Inc.\\
YHOO & Yahoo! Inc.\\
\noalign{\smallskip}\hline
\end{tabular}
\end{table}

\end{document}